\documentstyle[11pt]{article}

\begin{document}
\begin{center}
{\Large Effect due to compositeness of nucleons in deep inelastic
lepton nucleus scattering}

\vspace{3mm}
\renewcommand{\thefootnote}{\fnsymbol{footnote}}
{\normalsize Bo-Qiang Ma\footnote{
Fellow of Alexander von Humboldt Foundation.}
}

\vspace{3mm}
{\large Institute of High Energy Physics, Academia Sinica, P.O.Box 918(4),
Beijing 100039, China }

{\large Institut f\"ur Theoretische Physik der
        Universit\"at Frankfurt am Main, Postfach 11 19 32,
        D-6000 Frankfurt, Germany}

\vspace{2mm}
\end{center}
{\large \bf Abstract }
The off-shell behaviors of bound nucleons in deep inelastic lepton
nucleus scattering are discussed in two scenarios with the basic
constituents chosen to be baryon-mesons and quark-gluons respectively
in light-cone formalism. It is found that when taking into account
the effect due to internal quark structure of nucleons, the
derived scaling variable for bound nucleons and the calculated
nuclear structure functions are different from those in considering
the baryon-mesons as the effective elementary constituents. This
implies that the pure baryon-meson
descriptions of nuclei give the inaccurate off-shell
behavior of the bound nucleon structure function, thereby the
quark-gluons seem to be the most appropriate degrees of freedom for
nuclear descriptions.
It is also shown that the EMC effect cannot
be explained by nuclear binding effect from a sound theoretical
basis.

\vspace{8mm}
\break
\renewcommand{\theequation}{\thesection.\arabic{equation}}

\section{Introduction}

What are the most appropriate degrees of freedom for
nuclear descriptions? Can we develop a suitable effective quantum
field theory of the baryon-meson that can accurately describe the
detailed properties of the ordinary matter, or must we study the
nuclear structure and dynamics in terms of the underlying quark-gluon
component structure of hadrons? These questions are often asked by
nuclear and particle physicists in recent years. One of the commonly
accepted standpoints seems to be that nuclear
physics on the microscopic level can be consistently described using
nucleons, isobars and pions with some vector-meson contributions.
In the low energy region where the constituent baryon-mesons can
be considered as pointlike particles, this picture has been proved
to be extremely successful.
However, this picture may need improvement in the medium-and-high
energy region where the compositeness of the hadrons should not be
ignored further. In fact, it has been indicated by Brodsky several years
ago [1] that the effective local Lagrangian field theory
of the baryon-meson gives an inaccurate description of the actual
dynamical and off-shell behaviors of hadronic amplitudes due to
internal quark and gluon structure of the hadrons, hence
it is not predictive and reliable.
This viewpoint was stressed further recently
by Jaroszewicz and Brodsky [2] from comparing the Z-diagrams in two
pictures where the bound nucleons are treated as elementary Dirac
particles and composite quark objects respectively.
Jaffe also indicated [3] the limitations of the validity of effective field
theory of mesons and baryons from comparing the calculated hadron properties
from dispersion theory and those from quark model approaches.

This paper aims to emphasize Brodsky's viewpoint from the discussions
of the off-shell behaviors for bound nucleons in deep
inelastic lepton
nucleus scattering, i.e., the
nuclear European Muon Collaboration (EMC) effect [4]. We
will treat the scattering process in two scenarios with the
basic constituents chosen to be baryon-mesons
and quark-gluons respectively based on the
light-cone quantum field theory [5]. The first scenario is to consider
the nucleus as a bound state of baryon-mesons which are treated as
effective elementary Dirac particles with the structure
functions identified with those for free ones by a seemingly
natural kinematic analogy. It is shown that the EMC effect could be
phenomenally
very well reproduced by using minimal multihadron Fock state
(i.e., nucleons only) for certain $Q^{2}$ in taking into account
the off-"'energy''-shellness of bound nucleons.
However,
the ambiguities in identifying the structure functions for
bound nucleons seem to have not been avoided because there are ambiguities
in evaluating the off-"'energy''-shell effect in this scenario.
In order to avoid the off-"'energy''-shell ambiguities,
we consider, in the second scenario, the nucleus
as a bound state of baryon-mesons which are also composite
states of quark-gluons. The scaling variable for bound hadrons
is derived in a realistic way from the constraint of overall
"'energy''
conservation.  It is found that this scaling variable
for bound hadrons is different from that in the first scenario, hence
we obtain different results for the nuclear structure function.
The EMC effect cannot be explained by using minimal Fock state alone in
the latter case if we do not introduce the "'intrinsic'' distortions
in the structure functions for bound nucleons caused by nuclear
environment. Because the quark-gluon two-level convolution scenario
is more reasonable in taking
into account the correct dynamical and off-"'energy''-shell behaviors of
the structure function for bound nucleons, the calculated nuclear
structure function is more reliable than that in the pure baryon-meson
scenario. Thereby this work indicates
clearly that the compositeness of nucleons due to their internal
quark-gluon structure
should be taken into account
in the study of high momentum transfer processes such as the nuclear
EMC effect.

In Sect.2, the light-cone hadronic approach and the off-"'energy''-shell
ambiguities are illustrated. The formulas of the quark-gluon
two-level convolution scenario and some discussions
are presented in Sect.3. Sect.4 gives the summary and conclusions.

\section{The light-cone hadronic approach}

The EMC effect [4] is the fact that the measured
ratios of the structure functions for bound nucleons to those for free ones
are different from the earlier theoretical expectations [6] based on the
impulse approximation in the rest frame of the nucleus. This effect
was interpreted as a quark signature in nuclei when
it was discovered[7].
Many models were proposed to explain the observed data
by introducing non-nucleonic degrees of freedom such as nucleon swelling
or overlap, multi-quark cluster, color conductivity,
meson or isobar degrees
of freedom, and quark exchange between nucleons etc.
However, the situation has been
more or less blurred since
the appearance of a conventional nucleonic model [8] in which the EMC effect
can be interpreted in terms of nucleonic degrees of freedom alone if
the Fermi motion of nucleons and the nuclear
binding effect (i.e., off-mass-shell effect)
are taken into account. Because it was recognized
that there are ambiguities
in identifying the off-shell structure functions with the on-shell ones
[9-10]
and there is no a priori justification for ignoring final state interactions
[11-12] in the conventional model, an alternative model in solving the above
disadvantages simultaneously is obviously of importance for a further
insight view of the EMC effect.

It was deemed, in several light-cone nucleonic models [12-14] which adopt
the on-mass-shell kinematics for the bound nucleons, that the
ambiguities in identifying the off-mass-shell
structure function for bound nucleons with the on-mass-shell structure
function for free ones are avoided. Furthermore,
it has been indicated in ref.[12]
that final state interactions can be ignored by using instant
form dynamics in the infinite momentum frame,
or using light-front form dynamics in an ordinary frame.
Hence one may expect that the above two problems in conventional
nucleonic approach can be avoided by using light-cone hadronic
approaches
in dealing with deep inelastic lepton nucleus scattering.
In the following, we will exam whether this is the true case by discussing
the off-shell behaviors in the pure baryon-meson
scenario based on the light-cone quantum field theory of
baryon-meson fields.

\subsection{The pure baryon-meson scenario}

A relativistic composite model of nuclei in analogy with the
relativistic composite model of hadrons developed by Brodsky et al. [5]
has been proposed in ref.[12] by extending the light-cone quantum field
theory to baryon-meson fields. This model has the advantage that the
constituent baryon-mesons can be also visualized as composite
systems of quark-gluons by employing the
relativistic composite model of hadrons, with many of the nuclear properties
in conventional nuclear physics still being detained.
In this section, however,
we will not introduce the internal quark structure of the baryon-mesons.
We treat, instead, the baryon-mesons as the effective elementary constituents
of the nucleus and use their off-shell structure functions to account for
their compositeness.

It is adequate to use the minimal
multihadron Fock state (i.e., nucleons only) in the discussions of
the following sections because the main purpose
of this paper is to investigate the role played by
nucleonic degrees of freedom alone in the explanation
of the EMC effect. By applying the impulse approximation, which
has been justified in the light-cone approach[12],
we can illustrate the contribution to the hadronic tensor
$W^{A}_{\mu\nu}$ for the target nucleus in Fig.1, where the dotted
line means to sum over all possible final states. The kinematics
for the particles are parametrized as
\begin{equation}
\begin{array}{clcr}
P_{\mu}=(P^{+},P^{-},\vec{P}_{\perp})=(M_{A},M_{A},\vec{0}_{\perp});\\
q_{\mu}=(q^{+},q^{-},\vec{q}_{\perp})=(0,2\nu,\vec{q}_{\perp});\\
p_{\mu}=(p^{+},p^{-},\vec{p}_{\perp})=(yP^{+},(M^{2}+
\vec{p}_{\perp}^{2})/yP^{+},\vec{p}_{\perp}),
\end{array}
\end{equation}
provided with the defining equation
\renewcommand{\theequation}{\thesection.\arabic{equation}\ '}
\setcounter{equation}{0}
\begin{equation}
q^{2}=-Q^{2};\;\;\;   P\cdot q=M_{A}\nu.
\end{equation}
\renewcommand{\theequation}{\thesection.\arabic{equation}}
Using the calculation rules in ref.[12], we obtain
\begin{equation}
W^{A}_{\mu\nu}=\sum_{BM} \int [d^{2}\vec{p}_{\perp}dp^{+}/16\pi^{3}p^{+}]
[\rho_{BM}(\underline{p})/y]W^{N}_{\mu\nu}(p,p'),
\end{equation}
in which the function $\rho_{BM}(\underline{p})$ is the
light-cone momentum (i.e., $\underline{p}=(p^{+},\vec{p}_{\perp})$
distribution of nucleons in the nuclear bound state, and
$W^{N}_{\mu\nu}(p,p')$
is the hadronic tensor for the on-mass-shell struck nucleon
with its kinematics before and after the scattering being  $p_{\mu}$
and $p'_{\mu}$
respectively, where $p'_{\mu}$ subjects to the
constraint of overall 4-momentum
conservation between the virtual photon and the target
nucleus. Then we have
\begin{equation}
\begin{array}{clcr}
q'^{-}=p'^{-}-p^{-}=q^{-}+P^{-}_{A}-P^{-}_{C}-p^{-}
	=q^{-}-(\frac{\vec{p}_{\perp}^{2}+M_{C}^{2}}{M_{A}-p^{+}}
       + \frac{\vec{p}_{\perp}^{2}+M^{2}}{p^{+}}-M_{A}),
\end{array}
\end{equation}
where $M_{C}=\varepsilon+M_{A}-M$ is the mass of the residual
nucleus and $\varepsilon=M_{C}+M-M_{A}$
is the separation energy of a nucleon from the nucleus.

In ref.[12], the nuclear structure functions are obtained from
the calculation of $L^{\mu\nu}W^{A}_{\mu\nu}$ in which
the Z-graph contributions are
neglected.
In order to avoid the neglect of the Z-graph contributions,
we calculate only the ++ component of $W^{A}_{\mu\nu}$ because
there are no Z-graph contributions in it. Hence we obtain
the convolution formula for the nuclear structure function
\begin{equation}
F_{2}^{A}(\nu,q^{2})=\sum_{BM} \int [\frac{d^{2}\vec{p}_{\perp}dp^{+}}
{16\pi^{3}p^{+}}]\frac{M_{A}\nu}{p\cdot q}y\,\rho_{BM}(\underline{p})
F_{2}^{N}(p,p',q),
\end{equation}
in which $F_{2}^{N}(p,p',q)$ is the structure function for the struck
nucleon with $p_{\mu}$ and $p'_{\mu}$ being the kinematics before
and after the scattering,
and $q_{\mu}$ being the 4-momentum of the incident virtual photon.
In a strict sense,
the valve of $F^{N}_{2}(p,p',q)$ is not the same as that for a free
nucleon because the
kinematics are different for the two cases. This arises the problem
as to how to identify
the off-shell structure functions for bound nucleons
with those for free ones.
\subsection{The off-shell ambiguities}
\noindent
{\bf a) The on-shell case}

For a free nucleon at rest, we know its structure function can be expressed
as
\begin{equation}
F_{2}(p,p',q)=F_{2}(x,Q^{2})
\end{equation}
with $x=Q^{2}/2M\nu$ and $Q^{2}=-q^{2}$.
For a nucleon with velocity $\vec{v}$, its structure
function can be written as
\begin{equation}
F_{2}(p,p',q)=F_{2}(x',Q'^{2})
\end{equation}
with
\begin{equation}
x'=-q^{2}/2p\cdot q; \;\;\;  Q'^{2}=-q^{2}.
\end{equation}
Because the following relations hold for a free nucleon:
\begin{equation}
p^{2}=M^{2};\;\;\;  q^{2}=-Q^{2};\;\;\;  q_{\mu}=p'_{\mu}-p_{\mu},
\end{equation}
we may re-express $x'$, $Q'^{2}$ in different forms, such as by:
\begin {equation}
x'=Q'^{2}/2M\nu';\;\;\;   Q'^{2}=\nu'^{2}-\vec{q}\,'^{2}
\end{equation}
with $\nu'=\gamma (\nu-\vec{v}\cdot \vec{q})$,
$\vec{q}\,'=\gamma(\vec{q}-\nu\vec{v})$ and
$\gamma=(1-\vec{v}^{2})^{-1/2}$ as has been adopted
by Noble [15], and by:
\begin{equation}
x'=-q'^{2}/2p\cdot q'; \;\;\;   Q'^{2}=-q'^{2}
\end{equation}
with $q'_{\mu}=p'_{\mu}-p_{\mu}$ as will be adopted in this paper.
In fact, there in an infinity
$x'$ and $Q'^{2}$  with different expressions
but same values for free nucleons.

\noindent
{\bf b) The off-mass-shell ambiguities}

The conventional nucleonic approach [8] is evaluated in the Ferynman-Dyson
perturbation theory in the rest frame of the nucleus. Because 4-momentum
is conserved between the lepton and the struck nucleon, and the recoiling
(A-1) nucleus is on-mass-shell, the struck nucleon must be off-mass-shell
to ensure overall energy and momentum conservation. In this case, the
relation $q'_{\mu}=p'_{\mu}-p_{\mu}=q_{\mu}$
holds but the on-mass-shell relation $p^{2}=M^{2}$
does not hold; i.e., the struck nucleon is off-mass-shell. From overall
energy conservation we have $p_{\mu}=(M-\varepsilon,p)$.

In this case, eq.(2.10) equals to eq.(2.7) from
$q'_{\mu}=q_{\mu}$. However, eq.(2.9)
does not equal to eq.(2.7) as was indicated in ref.[15].
In ref.[8], $x'=-q^{2}/2p\cdot q$ is adopted
as the scaling variable for a bound nucleon and the EMC-SLAC data are able
to be reproduced by using nucleonic degrees of freedom alone in taking
into account the consequences from the off-mass-shellness. However, it
was argued by Noble in ref.[15] that $x'=Q'^{2}/2M\nu'$
should be adopted as the
scaling variable for a bound nucleon and that the non-nucleonic degrees of
freedom
are required for explaining the EMC effect by using this variable in the
convolution formula. The above two methods both seem to be reasonable as
illustrated by their users. This implies that
the off-mass-shell ambiguities
prevent reliable predictions of the off-mass-shell effect, and thereby the
conventional nucleonic approach has difficulty
in reliably evaluating the off-mass-shell
effect.

\noindent
{\bf c) The off-"'energy''-shell ambiguities}

The light-cone approach in this paper is based on light-cone perturbation
theory in which all particles are on-mass-shell, and $p^{+}$ and
$\vec{p}_{\perp}$ are conserved
at every vertex with $p^{-}$ not necessary conserved at every vertex.
Thereby
the relation $p^{2}=M^{2}$ holds but
$q'_{\mu}=q_{\mu}$ does not hold again. In this case the
off-mass-shell ambiguities are avoided, as has been claimed in some
previous literature. However, the off-shell ambiguities still exist
because the off-"'energy''-shell ambiguities replace the off-mass-shell
ambiguities now. From the calculation we know that the uses of
$x'=-q'^{2}/2p\cdot q'$
or $x'=-q^{2}/2p\cdot q$ as the scaling variables for bound nucleons will
give different results, as can be seen from Fig.2.
The scaling variable $x'=-q^{2}/2p\cdot q$
has been adopted in several light-cone
hadronic approaches [14,16] recently and it was found that the EMC effect
cannot be reproduced by the nucleonic degrees of freedom alone.
However, it was argued in ref.[17] that the use of
$x'=-q'^{2}/2p\cdot q'$
seems to be more natural than the use of
$x'=-q^{2}/2p\cdot q$ because the off-"'energy''-shell
effect seems to have been considered. Thereby one may conclude from Fig.2
that the large part of the EMC effect could
be explained by contributions from
the nuclear Fermi motion effect and the nuclear binding (e.g., the
off-"'energy''-shell effect) in terms of nucleonic degrees of freedom alone
in the light-cone hadronic approach, as
was done in the conventional nucleonic approach. Since the off-shell
ambiguities have not been avoided in the light-cone hadronic approach,
its predictions of the off-"'energy''-shell effect are also unreliable as
in the case of the conventional nucleonic approach.

\section{The relativistic two-level convolution model}
\setcounter{equation}{0}

The discussions in the precede section show that the off-shell ambiguities
in identifying the scaling variable for a bound nucleon seem to have
not been avoided in the light-cone hadronic approach.
This can explain why there are prescription dependence in the calculated
off-shell effect in nucleonic models, as was pointed out by
Kisslinger and Johnson recently[16].
We know that
the scaling variable for a free nucleon can be derived from the
quark-parton model of hadrons [18,5,19]. Hence we may expect that the scaling
variable for a bound nucleon can be also derived from the
relativistic composite model of nuclei if the quark-gluon component
structure of the hadrons are introduced. This inspired the author
to have developed a relativistic two-level convolution model for nuclear
structure functions [20] with the scaling variable for a bound nucleon
derived in a realistic way other than from assumptions.

\subsection{The quark-gluon two-level convolution scenario}

The scaling variable for a free nucleon has been derived in ref.[19]
in the relativistic composite model of hadrons with
the off-"'energy''-shell
effect (i.e., the overall energy conservation effect called in ref.[19])
of the
struck quark also considered. The calculation gave a good precocious
scaling variable $x_{p}$, which reduces to the Weizmann variable,
the Bloom-Gilman
variable and the Bjorken variable at some approximations respectively.
We now apply the technique in ref.[19] to derive the scaling variable for
a bound hadron.

We introduce the quark-gluon component structure of the hadrons into
the relativistic composite model of nuclei. In this scenario
the target nucleus is considered
as a composite system of hadrons which are also composite systems of
quark-gluons.  In neglecting quark interference corresponding to the quark
exchange contributions, we can illustrate the contributions
to the hadronic tensor
$W^{A}_{\mu\nu}$ for the target nucleus in Fig.3,
where the reference frame is the
same one in the first scenario and the kinematics for
the quarks are parametrized as
\begin{equation}
k_{\mu}=(k^{+},k^{-},\vec{k}_{\perp})=(xp^{+},(m^{2}+
\vec{k}^{2}_{\perp})/xp^{+},\vec{k}_{\perp})
\end{equation}
Similar to the first scenario, we obtain
\begin{equation}
W^{A}_{\mu\nu}=\sum_{BM} \int [\frac{d^{2}\vec{p}_{\perp}dp^{+}}
{16\pi^{3}p^{+}}]\frac{\rho_{BM}(\underline{p})}{y}
\sum_{q}\int[\frac{d^{2}\vec{k}_{\perp}dk^{+}}
{16\pi^{3}k^{+}}]\frac{\rho_{BM}(\underline{k})}{x}w_{\mu\nu}(k,k')
\end{equation}
in which  $\rho_{BM}(\underline{p})$ is the same one used
in the first scenario, and
\begin{equation}
\begin{array}{clcr}
\rho_{q}(\underline{k})=\int\prod_{i=2}^{n}[\frac{d^{2}\vec{k}_{i\perp}
dk^{+}_{i}}{16\pi^{3}k^{+}_{i}}]16\pi^{3}\delta^{2}
(\vec{p}_{\perp}-\vec{k}_{\perp}-\sum_{i=2}^{n}\vec{k}_{i\perp}) \\
\delta(1-x-\sum_{i=2}^{n}x_{i})
|\psi_{q}(\underline{p};\underline{k},\underline{k}_{2},
\cdots \underline{k}_{n})|^{2};
\end{array}
\end{equation}
\begin{equation}
\begin{array}{clcr}
w_{\mu\nu}(k,k')=2Q_{q}^{2}[k_{\mu}k'_{\nu}+k_{\nu}k'_{\mu}
-g_{\mu\nu}(k\cdot k'-m^{2})] \\
\delta(k'^{-}+\sum_{j=2}^{m}k_{j}^{-}
+\sum_{i=2}^{l}p_{i}^{-}-P^{-}-q^{-})/k'^{+};
\end{array}
\end{equation}
The $x$, $y$ in the denominator of (3.2) are essential corresponding
to the flux factors in ref.[21-23] which guarantee baryon
number conservation,
$\rho_{q}(\underline{k})$ is the quark momentum distribution in
the struck baryon or meson, and $w_{\mu\nu}(k,k')$
is the hadronic tensor of the
on-mass-shell struck quark with its kinematics before and after
the scattering being $k_{\mu}$ and $k'_{\mu}$
respectively, supplying $k'_{\mu}$ subjects
to the constraint of overall 4-momentum conservation between the
virtual photon and the target nucleus[19]:
\begin{equation}
\begin{array}{clcr}
\vec{k}'_{\perp}=\vec{q}_{\perp}+\vec{k}_{\perp};\\
k'^{+}=q^{+}+k^{+};\\
k'^{-}+\sum_{j=2}^{m}k_{j}^{-}+\sum_{i=2}^{l}p_{i}^{-}=P^{-}+q^{-}.
\end{array}
\end{equation}
To obtain $F^{A}_{2}=\nu W^{A}_{2}$,
we calculate only the ++ component of $W^{A}_{\mu\nu}$ since the
instantaneous fermion lines do not contribute to it as have been
indicated by Brodsky et al.[5].
Using the relation
\begin{equation}
      F^{T}_{2}=(p_{T}\cdot q/2p^{+}_{T}p^{+}_{T})W_{++}^{T},
\;\;\; T=A,N,
\end{equation}
we obtain the two level convolution formula
\begin{equation}
F_{2}^{A}(x,Q^{2})=\sum_{BM} \int [\frac{d^{2}\vec{p}_{\perp}dp^{+}}
{16\pi^{3}p^{+}}]\frac{M_{A}\nu}{p\cdot q}y\,\rho_{BM}(\underline{p})
F_{2}^{N}(x_{B},Q^{2}),
\end{equation}
where
\begin{equation}
F_{2}^{N}(x_{B},Q^{2})=\sum_{q}\int dx\delta (x-x_{B})xQ^{2}_{q}
f_{q}(x,Q^{2})\frac{q^{-}}{q^{-}+k^{-}},
\end{equation}
in which $f_{q}(x)$ is the quark distribution and $x_{B}$ is the scaling
variable for the bound hadron. The factor $q^{-}/(q^{-}+k^{-})$ equals to
$1$ in the Bjorken limit. $x_{B}$ is obtained
from the overall "'energy'' conservation condition[19]
\begin{equation}
\frac{[m^{2}+(\vec{k}_{\perp}+\vec{q}_{\perp})^{2}]}
{(k^{+}+q^{+})}+\sum_{j=2}^{m}k_{j}^{-}+\sum_{i=2}^{l}p_{i}^{-}
=\frac{(M^{2}_{A}+\vec{P}^{2}_{\perp})}{P^{+}}+q^{-}.
\end{equation}
Replacing $\sum_{j=2}^{m}k_{j}^{-}$ and $\sum_{i=2}^{l}p_{i}^{-}$
by $p_{c}^{-}$ and $P_{C}^{-}$,  the  minus  component
momentums of the residual hadron, nucleus respectively, we have
\begin{equation}
x_{B}=(A-B)/2C,
\end{equation}
in which
\setcounter{equation}{9}
\renewcommand{\theequation}{\thesection.\arabic{equation}\ '}
\begin{equation}
\begin{array}{clcr}
    A=C+m^{2}+(\vec{k}_{\perp}+\vec{q}_{\perp})^2-m_{c}^{2}
-(\vec{p}_{\perp}-\vec{k}_{\perp})^{2};\\
B=\{A^{2}-4C[m^{2}+(\vec{k}_{\perp}+\vec{q}_{\perp})^{2}]\}^{1/2};\\
    C=[M^{2}_{A}+2M_{A}\nu-(\vec{p}_{\perp}^{2}+M^{2}_{C})/(1-y)]/y.
\end{array}
\end{equation}
\renewcommand{\theequation}{\thesection.\arabic{equation}}
In the Bjorken limit $Q^{2}\rightarrow\infty$, $\nu\rightarrow\infty$
with $x=Q^{2}/2M\nu$ fixed, $x_{B}$ reduces to
\begin{equation}
       x_{B}=Q^{2}/2M_{A}\nu y.
\end{equation}
When comparing the above results with the calculation of the structure
function for a free hadron in ref.[19],
one sees that there are no ambiguities in
identifying the structure functions for bound hadrons with those
for free ones in assuming that there are no "'intrinsic'' distortions
in the quark distribution for bound hadrons caused by
nuclear environment (e.g.,
nucleon swelling). However, the scaling variable for a bound hadron
(i.e., $x_{B}$ in this paper) is different from that for a free hadron
($x_{p}$ in ref.[19]) because the bound hadron is off-"'energy''-shell
and subjects to the overall "'energy'' conservation.
In this sense, the binding effect (i.e., off-"'energy''-shell effect)
is also included in the two level
convolution formula (3.7) from the constraint of
overall "'energy'' conservation. The
contributions from nuclear binding are contained in
$M_{C}$ in the scaling
variable $x_{B}$, it gives $Q^{2}$
power-law type contributions and hence can be
neglected if $Q^{2}$ and $\nu$ are sufficient large [19].
It becomes clear that the
above result is different from that in the first scenario which gives a
scaling variable $x'=-q'^{2}/2p\cdot q'$ for a bound nucleon.

\subsection{The calculated results and discussions}

Fig.4 presents the calculated ratios
$F^{A}_{2}(x,q^{2})/F^{N}_{2}(x,q^{2})$ in the
quark-gluon two-level convolution scenario with
the necessary inputs being the same as those used in ref.[12].
It is found that the results are different from those in
the pure baryon-meson scenario. In the pure baryon-meson
scenario, the large part of the EMC effect is
able to be reproduced
very well for $Q^{2}=5$ (GeV/c)$^{2}$;
whereas in the quark-gluon two-level convolution scenario
the calculation cannot reproduce
the data by using nucleonic degrees of freedom alone. The results
in the quark-gluon two-level convolution scenario
should be more reliable because the
effect due to internal quark structure of nucleons is explicitly considered.
Hence the results indicate that the EMC effect cannot
be explained by nucleonic degrees of freedom alone if one properly
considers the off-shell behaviors
of nucleons due to their compositeness.

The difference between the baryon-meson and the quark-gluon
scenarios can be seen from eq.(2.4)
and eq.(3.7): The scaling variables for bound nucleons are different
in the two scenarios though the kinematics for the struck nucleon are
the same. In the pure baryon-meson scenario, the scaling
variable for bound nucleons ($x'=-q'^{2}/2p\cdot q'$)
is identified with that
for free ones by a seemingly most natural kinematic analogy.
Whereas in the quark-gluon
two-level convolution scenario, the scaling variable for the bound nucleon
(i.e., eq.(3.10)) is derived in a realistic way from the constraint
of overall "'energy'' conservation; i.e., in a way similar to that
in which the Bjorken scaling variable was derived in the quark-parton
model[18,5,19]. The difference hence indicates that the local Lagrangian
field theory of the baryon-meson gives the inaccurate description
of the actual dynamical and off-"'energy''-shell behaviors of the
structure function for bound nucleons due to their internal quark-gluon
component structure.

We indicate here that the above conclusion does not rely on the
specific form of the light-cone momentum distribution function
$\rho_{BM}(\underline{p})$ used in this paper.
The factor $(M_{A}\nu/p\cdot q)y$ in the two-level convolution
formula eq.(3.7) equals to unity for large $Q^{2}$ and $\nu$.
Glazek and Schaden have
adopted a parton-like convolution formula, which is equivalent to
eq.(3.7) for large $Q^{2}$ and $\nu$,
to investigate the role of pionic degrees of
freedom in the EMC-SLAC data [24]. From their analysis we know that the
ratio $r$ will be equal to $1$ around $x=0.6$,
where the data show large
deviation from $1$, if the nucleonic degrees of freedom alone are considered.
This implies that non-nucleonic degrees of freedom are unambiguously
required in the explanation of the EMC effect.

\section{Summary and conclusions}

The off-shell behaviors of bound nucleons
in deep inelastic lepton nucleus scattering are discussed in two
scenarios based on light-cone perturbation theory in this
paper. It is shown that the off-"'energy''-shell effect is not
reliably handled in the pure baryon-meson scenario though
the results may be good in describing the data (for discussions
of some detailed features in the light-cone hadronic scenario,
see ref.[17]).
When taking into account the effect due to internal
quark structure of hadrons, the EMC effect cannot be reproduced by
using nucleonic degrees of freedom alone.
This indicates that the nucleonic approaches
have also the disadvantage that the effects due to compositeness
of bound nucleons are unable to be reliably handled. Thereby after several
years in elusiveness caused by the conventional nucleonic approach,
the former conclusion [7] that
the EMC unambiguously indicates the presence of non-nucleonic
degrees of freedom seems to have been recovered.
The inclusion of isobar and meson
degrees of freedom (i.e., higher multihadron Fock state components),
or the introduction of some
other exotic effects
such as the intrinsic distortions of the quark momentum distribution
for hadrons caused by nuclear environment, quark exchange between
nucleons (or multiquark cluster),  color conductivity,
and hidden color component et al. are required in explaining
the EMC effect.

One conclusion from this work is that it is probably inappropriate to
treat the nucleus as a composite system of hadrons without considering
their underlying quark-gluon structure in the medium and
high energy physics regions. Hence much work should be done
to analyze the effects due to compositeness of hadrons
in various reactions and nuclear properties though
they are not easy tasks. The questions asked in the beginning of
this paper seem to have been answered by this work: The quark-gluons
are essentially the most appropriate degrees of freedom for nuclear
descriptions though the baryon-mesons are also the relevant
degrees of freedom in nuclear physics.

\noindent
{\large \bf Acknowledgement}

The author would like to thank Prof.Tao Huang and Prof.Ji Sun for
valuable discussions, and to acknowledge the encouraging discussions
with Prof.M.B.Johnson.  He is also grateful to Prof.W.Greiner for his
hospitality and for the support from Institut f\"ur Theoretische
Physik der Universit\"at Frankfurt.

\break
\noindent
{\large \bf References}
\begin{enumerate}

\item
S.J.Brodsky, Comments Nucl.Part.Phys.12(1984)213;
    in Short-Distance Phenomena in Nuclear Physica, eds.
    David H.Boal and Richard M.Woloshyn (Plenum, New York, 1983) p.141.
\item
T.Jaroszewicz and S.J.Brodsky, Phys.Rev.C43(1991)1946.
\item
R.L.Jaffe, Nucl.Phys.A522(1991)365c.
\item
EMC, J.J.Aubert, et al., Phys.Lett.123B(1983)275;

    A.Bodek, et al., Phys.Rev.Lett.50(1983)1431 and 51(1983)534.

    For new EMC data, see, EMC, J.Ashman, et al., Phys.Lett.202B(1988)603.
\item
S.J.Brodsky, in Lectures on Lepton Nucleon Scattering and
    Quantum Chromodynamics, eds. A.Jaffe and D.Ruelle
    (Birkh\"auser, Boston, 1982) p.255;

    S.J.Brodsky, T.Huang, and G.P.Lepage, in Particles and Fields,
    eds. A.Z.Capri and A.N.Kamal (Plenum, New York, 1983) p.143.
\item
A.Bodek and J.L.Ritchie, Phys.Rev.D23(1981)1070; D24(1981)1400.
\item
R.L.Jaffe, Phys.Rev.Lett.50(1983)228.
\item
S.V.Akulinichev, et al., Phys.Rev.Lett.55(1985)2239;
    J.Phys.G11(1985) L245; Phys.Lett.158B(1985)485.

    See also, e.g., B.L.Birbrair, et al., ibid.166B(1986)119;
    G.V.Dunne and A.W.Thomas, Nucl.Phys.A446(1985)437c;
    Phys.Rev.D33(1986)2061; Nucl.Phys.A455(1986)701.
\item
M.B.Johnson and J.Speth, Nucl.Phys.A470(1987)488;

    B.-Q.Ma and J.Sun, High Energ.Phys.Nucl.Phys.12(1988)337;

    L.Heller and A.W.Thomas, Phys.Rev.C41(1990)2756.
\item
A.W.Thomas, Prog.Part.Nucl.Phys.20(1987)21.

\item
R.L.Jaffe, Comments Nucl.Part.Phys.13(1984)39; in Relativistic
    Dynamics and Quark-Nuclear Physics, eds. M.B.Johnson and
    A.Picklesimer (Wiley-Interscience, New York, 1986) p.71.
\item
B.-Q.Ma and J.Sun, J.Phys.G:Nucl.Part.Phys.16(1990)823.
\item
E.L.Berger and F.Coester, in Workshop on Nuclear Chromodynamics:
    Quarks and Gluons in Particles and Nuclei, eds. S.J.Brodsky
    and E.J.Moniz, (World Scientific, Singapore, 1986) p.255;

    See also, E.L.Berger, F.Coester, and R.B.Wiringa, Phys.Rev.D29(1984) 398;
    E.L.Berger and F.Coester, ibid.D32(1985)1071.
\item
U.Oelfke, P.U.Sauer, and F.Coester, Nucl.Phys.A518(1990)593.
\item
J.V.Nobel, in Symmetry Violations in Subatomic Physics, eds.
    by B.Castel and P.J.O'Donnell (World Scientific, Singapore, 1989) p.161.
\item
L.S.Kisslinger and M.B.Johnson, Phys.Lett.B259(1991)416.
\item
B.-Q.Ma, Mod.Phys.Lett.A6(1991)21.
\item
S.D.Drell, D.J.Levy, and T.-M.Yan, Phys.Rev.187(1969)2159;
    ibid.D1 (1970)1035;

    S.D.Drell and T.-M.Yan, Ann.Phys.(N.Y.)66(1971)578.
\item
B.-Q. Ma, Phys.Lett.176B(1986)179;

    B.-Q. Ma and J.Sun, Int.J.Mod.Phys.A6(1991)345.
\item
B.-Q.Ma, Phys.Rev.C43(1991)2821.
\item
L.L.Frankfurt and M.I.Strikman, Phys.Lett.183B(1987)254;
    Phys.Rep. 160(1988)235.
\item
G.L.Li, K.F.Liu, and G.E.Brown, Phys.Lett.213B(1988)531.
\item
C.Ciofi degli Atti and S.Liuti, Phys.Lett.225B(1989)215.

    For similar discussions, see,
    S.Shlomo and G.M.Vagradov, Phys.Lett.232B (1989)19;
    S.V.Akulinichev and S.Shlomo, ibid.234B(1990)170.
\item
St.Glazek and M.Schaden, Z.Phys.A323(1986)451.
\end{enumerate}

\break
\noindent
{\large \bf Figure Captions}
\renewcommand{\theenumi}{\ Fig.\arabic{enumi}}
\begin{enumerate}
\item
    The contributions to the hadronic tensor $W^{A}_{\mu\nu}$
    for the nucleus in
    the relativistic pure baryon-meson scenario.
\item
    The calculated ratio $F^{A}_{2}(x,Q^{2})/F^{N}_{2}(x,Q^{2})$
    in the pure baryon-meson
    scenario in comparison with the
    data. The solid and dashed curves are the results
    with $q'_{\mu}\not=q_{\mu}$ and $q'_{\mu}=q_{\mu}$
    respectively for $k_{F}=260$MeV, $Q^{2}=5$(GeV/c)$^{2}$ and
    $\varepsilon=35$MeV.
    The $\;\;\;$ points are the SLAC data for A=Fe ($Q^{2}=5$(GeV/c)$^{2}$,
    $8<\epsilon<24.5$GeV)
    and the $\;\;\;$ points are the new EMC data for A=Cu
    ($4.4$(GeV/c)$^{2}<Q^{2}<40.4$(GeV/c)$^{2}$,
    $\epsilon=120-280$GeV) from ref.[4] respectively.
\item
    The contributions to the hadronic tensor $W^{A}_{\mu\nu}$
    for the nucleus in
    the relativistic quark-gluon two-level convolution scenario.
\item
    The calculated ratio $F^{A}_{2}(x,Q^{2})/F^{N}_{2}(x,Q^{2})$
    in the quark-gluon
    two-level convolution scenario in comparison with the
    data. The solid curve are the results in  the quark-gluon two-level
    convolution scenario for $k_{F}=260$MeV, $Q^{2}=5$(GeV/c)$^{2}$
    and $\varepsilon=35$MeV
    with the dashed curve are the results
    in the pure baryon-meson scenario for comparison.

\end{enumerate}

\end{document}